\documentclass[runningheads]{llncs}
\usepackage{graphicx}
\usepackage{subcaption}
\usepackage{xcolor}
\usepackage{multirow}
\usepackage{booktabs}
\usepackage{url}
\usepackage{xspace}
\newcommand\surajnotes[1]{}
\newcommand\eugenenotes[1]{}
\newcommand\dawnnotes[1]{}
\newcommand\kevinnotes[1]{}
\newcommand\paulnotes[1]{}
\newcommand\kentonnotes[1]{}
\newcommand\jimnotes[1]{}
\newcommand\dougnotes[1]{}

\newcommand\colbertx{ColBERT-X\xspace}
\newcommand\xlmr{XLM-R\xspace}

\begin{document}

\title{Transfer Learning Approaches for Building Cross-Language Dense Retrieval Models}
\titlerunning{Cross-Language Dense Retrieval}
%
 \author{Suraj Nair\inst{1,2}\orcidID{0000-0003-2283-7672} \and
 Eugene Yang\inst{2}\orcidID{0000-0002-0051-1535} \and \\
 Dawn Lawrie\inst{2}\orcidID{0000-0001-7347-7086} \and 
 Kevin Duh\inst{2}\orcidID{0000-0001-8107-4383} \and \\
 Paul McNamee\inst{2}\orcidID{0000-0002-0548-5751} \and
 Kenton Murray\inst{2}\orcidID{0000-0002-5628-1003} \and
 James Mayfield\inst{2}\orcidID{0000-0003-3866-3013} \and
 Douglas W. Oard\inst{1,2}\orcidID{0000-0002-1696-0407}}

 \authorrunning{S. Nair et al.}
%
 \institute{University of Maryland, College Park MD 20742, USA  \email{\{srnair,oard\}@umd.edu}
\and
 HLTCOE. Johns Hopkins University, Baltimore  MD 21211, USA \\
 \email{\{eugene.yang,lawrie,mcnamee,kenton,mayfield\}@jhu.edu,kevinduh@cs.jhu.edu}}
\maketitle              
\begin{abstract}
The advent of transformer-based models such as BERT has led to the rise of neural ranking models.
These models have improved the effectiveness of retrieval systems well beyond that of lexical term matching models such as BM25.
While monolingual retrieval tasks have benefited from large-scale training collections such as MS MARCO 
and advances in neural architectures,
cross-language retrieval tasks have fallen behind these advancements. 
This paper introduces \colbertx,
a generalization of the ColBERT multi-representation dense retrieval model
that uses the XLM-RoBERTa (\xlmr) encoder to support cross-language information retrieval (CLIR).
\colbertx can be trained in two ways.
In \textit{zero-shot} training, the system is trained on the English MS MARCO collection,
relying on the \xlmr encoder for cross-language mappings.
In \textit{translate-train}, the system is trained on the MS MARCO English queries
coupled with machine translations of the associated MS MARCO passages.
Results on ad hoc document ranking tasks in several languages
demonstrate substantial and statistically significant improvements of these trained dense retrieval models
over traditional lexical CLIR baselines.

\end{abstract}

\section{Introduction}

BERT-style neural ranking models that use cross-attention between query and document terms~\cite{devlin-etal-2019-bert,liu2019roberta}
define the state of the art for monolingual English retrieval. 
Such models are typically used as rerankers in a retrieve-and-rerank pipeline,
due to the quadratic time and space complexity of self-attention in the transformer architecture~\cite{vaswani2017attention}. 
Reranking using these models is effective but time-consuming, so the number of documents to be reranked must be tuned to balance the trade-off between effectiveness and efficiency.  
In contrast to the reranking approach,
dense retrieval models encode query and document representations independently and match them with custom similarity functions (e.g., cosine similarity).
Dense retrieval complements the lexical first phase retrieval by using an approximate nearest neighbor search over contextualized representations.

While the retrieve-and-rerank framework has been adapted and explored in cross-language information retrieval (CLIR)~\cite{zhang2019improving,Jiang2020-dt,Zhao2019-el,Bonab2020-xk,Yu2020-dl},
most approaches translate queries into the language of the documents
and perform monolingual retrieval~\cite{shi2019cross,shi2021cross}. 
Dense retrieval models, on the other hand, remain under-explored in CLIR. 
In this work, we develop an effective dense retrieval model for CLIR. 

Dense retrieval models can be broadly categorized into two variants: single-representation and multi-representation~\cite{lin2021pretrained}.
Single-representation models encode queries and documents separately to create a single aggregate representation. 
However, that can lead to loss of information. 
Multi-representation models use multiple representations of queries and documents to predict relevance.
One such model is ColBERT~\cite{Khattab2020-ba},
which computes a similarity between each query term representation and each document term representation.
Yet ColBERT is exclusively monolingual.
This paper presents \textit{\colbertx}, a generalization of the ColBERT approach that supports CLIR. 
\colbertx uses a translate and train fine-tuning approach to exploit existing CLIR training resources. 

This generalization poses two challenges:
enabling the encoders to process multiple languages,
and identifying appropriate resources with which to train the model. 
To address the former, we adapt \xlmr~\cite{Conneau2019-sa},
a multilingual pretrained transformer language model,
to initialize the dense retrieval model.
 For the latter challenge, we use translations of MS MARCO~\cite{bajaj2018ms},
a widely-used passage ranking collection for training monolingual neural retrieval models.

We evaluate \colbertx on ad hoc document ranking tasks using English queries to retrieve documents in other languages, exploring
two ways to cross the language barrier.
In the zero-shot setting, where we lack cross-language training resources,
we train the model only on English MS MARCO.
In the translate-train setting, the model is trained on machine-generated translations of MS MARCO passages
paired with English queries. This paper additionally investigates the effect of machine translation on \colbertx retrieval results. 

Our main contributions can be summarized as follows:
\begin{itemize}
    \item We generalize ColBERT to support CLIR
    and develop a fine-tuning task that leverages translations of existing monolingual retrieval collections. 
    \item We demonstrate significant effectiveness gains
    over query translation baselines on news in several languages, showing the ability of term-level Approximate Nearest Neighbor (ANN) search to overcome vocabulary mismatch. 
    \item We analyze components of \colbertx and techniques to improve effectiveness,
    including effects of different machine translation models, alternative multilingual encoders, and relevance feedback.
    \item We release our code to train and evaluate \colbertx,
    and our new machine translations of MS MARCO into Chinese, Persian and Russian.\footnote{\url{https://github.com/hltcoe/ColBERT-X}}
\end{itemize}

\section{Related Work}

In this section, we briefly review related work on neural retrieval and its extension to cross-lingual settings. 
For many years, sparse retrieval models such as BM25~\cite{robertson1995okapi} and Query Likelihood~\cite{ponte1998language} were the dominant models for ad hoc retrieval tasks.
Only in recent years, with the rise of BERT~\cite{devlin-etal-2019-bert} and the availability of large scale retrieval collections
such as MSMARCO~\cite{bajaj2018ms} for training,
have neural information retrieval~(neural IR) models emerged as the state of the art.

Similar to sparse retrieval models,
neural IR models take as input the query and documents, and produce a relevance score.
For each query and document pair, matching mechanisms, such as DRMM~\cite{guo2016drmm}, KNRM~\cite{dai2018knrm} or PACCR~\cite{hui2017pacrr},
construct the interaction matrix between the distributed term representations of the query and the documents,
and aggregate them into a relevance score. 
Alternatively, the BERT passage pair classification model~\cite{devlin-etal-2019-bert} considers the query and the document as the input pair,
and uses the final classification score as the relevance score~\cite{yang2019vanilla-bert}. 
CEDR~\cite{macavaney2019cedr} incorporates contextualized embeddings such as ELMo~\cite{peters2018elmo} or BERT~\cite{devlin-etal-2019-bert}
into the matching,
providing significant effectiveness improvements by taking advantage of contextualization.
However, due to the high computational cost, these models are used to rerank top-ranked documents from a sparse retrieval system. 

ColBERT~\cite{Khattab2020-ba} further improves efficiency by keeping separate the query-document interaction until the end of the neural architecture.
This is called \textit{late interaction}.
As opposed to matching mechanisms that require both the query and the document to be present simultaneously,
late interaction allows offline encoding of the documents into bags-of-vectors.
Document representations are combined with query representations by an efficient \textit{MaxSim} operator,
which significantly reduces computation at inference time.
This decoupling enables the documents to be encoded offline
and indexed to support approximate nearest neighbor search.
Further details are discussed in Section~\ref{sec:methods}. 

Cross-language transfer learning is important for CLIR.
Due to the lack of training data for ad hoc neural retrieval models other than in English,
prior work explored zero-shot model transfer to other languages,
trained with only English retrieval examples~\cite{macavaney2020teaching,shi2019cross}.
Model initialization with a multilingual language model such as mBERT~\cite{devlin-etal-2019-bert} has been shown to be effective in zero-shot evaluations.
However, this approach requires both queries and documents to be in the same language,
resulting in evaluation based either on monolingual non-English retrieval~\cite{macavaney2020teaching},
or on query translation into the target language~\cite{shi2019cross}. 

With the availability of translations of the widely-used English ad hoc retrieval resource MS MARCO~\cite{bonifacio2021mmarco},
translate-train (training the retrieval model on a translated collection) using large ad hoc retrieval collections becomes feasible.
Prior work explored a dense retrieval approach to translate-train,
showing effectiveness gains on monolingual non-English retrieval tasks~\cite{shi2021cross}.
However, this approach relied on a single-representation dense retrieval model with an mBERT encoder, combined with sparse retrieval methods such as BM25. Lacking an end-to-end CLIR dense retrieval model that does not require the help of a sparse retrieval system,
we bridge the gap by generalizing ColBERT to support directly querying non-English documents with English queries.

\section{\colbertx}\label{sec:methods}

ColBERT is a multi-stage dense retrieval model that uses monolingual BERT~\cite{devlin-etal-2019-bert} to encode both query and document terms. It employs a late-interaction mechanism, MaxSim, that computes the similarity between the encoded query and document term representations. Computing MaxSim for every query and document term pair in the collection is not feasible, so ColBERT has two ways to reduce the number of required similarity comparisons: reranking or end-to-end retrieval.
In reranking, a retrieval system such as BM25 generates an initial ranked list, which is then reranked using ColBERT's MaxSim operation.
The disadvantage of such a cascaded pipeline is that the overall recall of the system is limited to the recall of the initial ranked list.
In the context of CLIR systems, we face the additional complexity of crossing the language barrier that further affects recall. We thus restricted our work to end-to-end (E2E) retrieval. 

In the first stage of the E2E setting, a candidate set of documents is generated by ANN search using every query term.
Specifically, the k nearest document tokens are retrieved from the ANN index for every query term representation.
These tokens are mapped to document IDs, and the union of these IDs creates the final set of candidate documents.
In the next stage, these documents are reranked using the late-interaction ``MaxSim'' operation.
For every query term, MaxSim finds the closest document token using the dot product of the encoded query and document term representation.
The final score of the document is the summation of individual query term contributions, as shown in Equation~\ref{eqn:colbert}.
$\eta$ denotes the monolingual BERT encoder.
\begin{equation}
    s_{q,d} = \sum_{i=1}^{|q|} \max_{j=1..|d|} \eta(q_i)\ .\ \eta(d_j)^{T} \label{eqn:colbert}
\end{equation}

To generalize ColBERT to CLIR, 
we replaced monolingual BERT with \xlmr. We call the resulting model \colbertx. Initializing the encoder to a multilingual model allows retrieval in any language supported by the embeddings. However, these models must be trained before they can be used for CLIR.

\subsection{CLIR Training Strategies}
\label{sec:ColBERT-X_training}
\begin{figure}[t!]
\centering
\includegraphics[width=\textwidth]{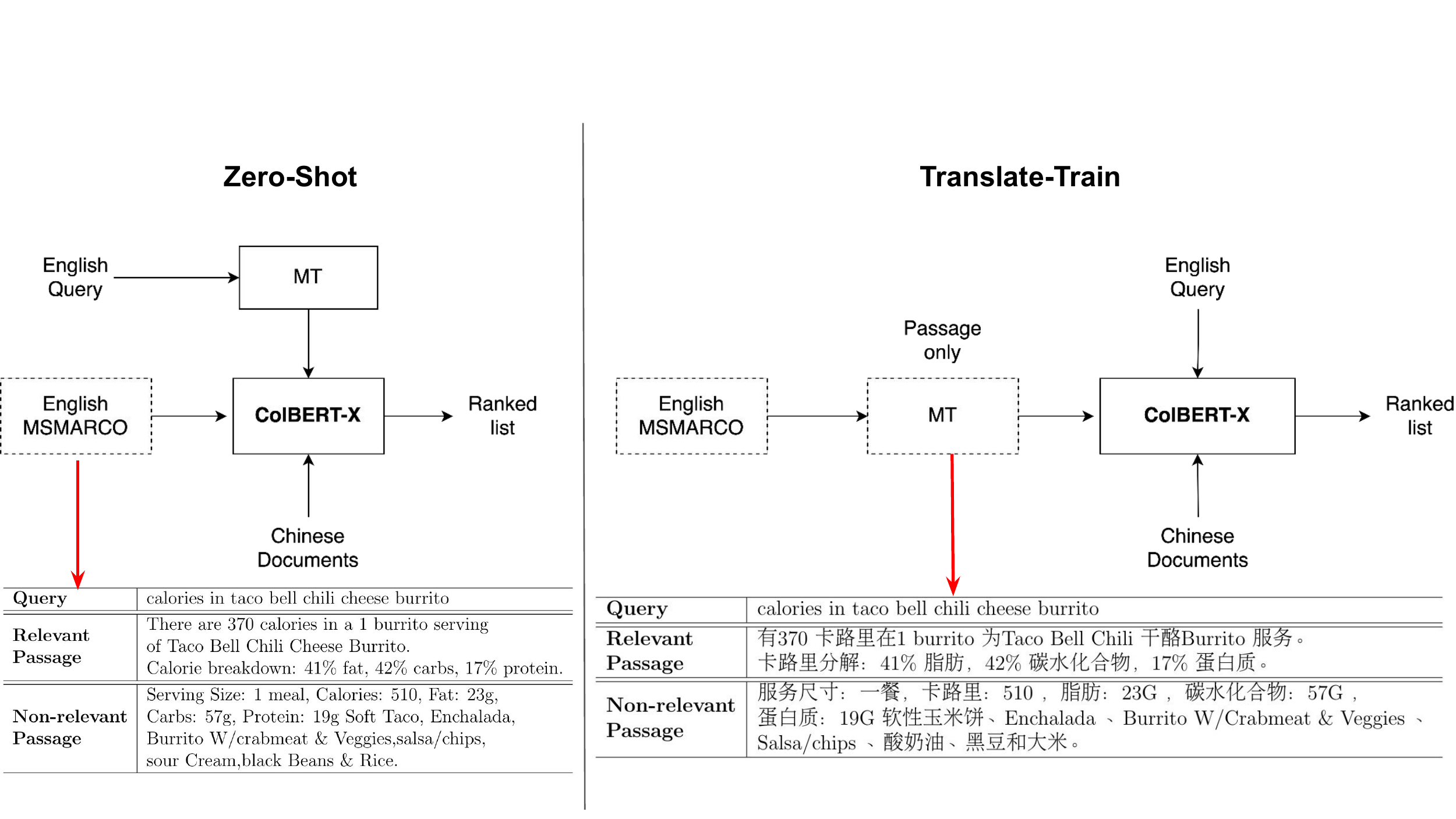}
\caption{\label{fig:colbertx_pipeline} Two \colbertx Transfer Learning Pipelines: Zero-Shot (left) and Translate-Train (right). Dashed boxes denote components used during the training step. In zero-shot, \colbertx trained on English MS MARCO is applied on the machine translated queries. With translate-train, the training set consists of translated passages to enable \colbertx to cross the language barrier.}
\end{figure}

ColBERT was trained using pairwise cross-entropy loss on MS MARCO~\cite{bajaj2018ms} triples,
which consist of an English query, a relevant English passage, and a non-relevant English passage.
To train \colbertx for CLIR, we explored two strategies from the cross-language transfer learning literature:
\begin{enumerate}
\item Zero-Shot: 
This is a common technique in which a multilingual model (e.g., mBERT or \xlmr)
is trained in a high-resource language (usually English)
and then applied to the document language.
In this paper, we first train a \colbertx model initialized with an \xlmr encoder
on English MS MARCO passage ranking triples.
At query time, we use machine translation (MT) to translate the English query to the document language,
and use the trained \colbertx model to perform retrieval in the document language using Equation \ref{eqn:colbertx-zs}.
$\hat{q}$ is the translated query.
Multilingual language models have demonstrated good cross-language generalization in many other natural language processing tasks;
we hypothesized it would also work well for CLIR.
\begin{equation}
    s_{\hat{q},d} = \sum_{i=1}^{|\hat{q}|} \max_{j=1..|d|} \eta(\hat{q}_i)*\eta(d_j) \label{eqn:colbertx-zs}
\end{equation}

\item Translate-Train: In this setting, an existing high-resource language (e.g., English) collection
is translated to the document language.
As in zero-shot training,
we choose training triples from the MS MARCO passage ranking collection and use a trained MT model to translate them.
Since our focus here is using English queries to retrieve content in non-English languages,
we pair the original English queries with machine translations of relevant and non-relevant MS MARCO passages to form new triples.\footnote{If we had wanted to experiment with using non-English queries to find English content, we could have instead translated only the MS MARCO queries.}
We then train \colbertx on these newly constructed triples in the same manner as ColBERT.
\end{enumerate}

Figure~\ref{fig:colbertx_pipeline} shows these two pipelines.
The key difference is that in the zero-shot setting we have a single \colbertx model for a given query language
(in this case English)
that is used for retrieval in multiple document languages.
In the translate-train setting, we train a \colbertx model for each query-document language pair.
We might also combine translations in multiple languages to train a single multilingual \colbertx model, but we leave that for future work.

\subsection{Retrieval}
While we train \colbertx on passages,
our goal is to rank documents.
We split large documents into overlapping passages of fixed length with a stride.
During indexing, we use the trained \colbertx model to generate term representations from these passages.
These representations are stored in a FAISS-based ANN index~\cite{JDH17}
and are saved to disk for subsequent MaxSim computation.
At query time, we use the trained \colbertx model to generate a ranked list of passages for each query using the approaches discussed in the section above and then use a document's maximum passage score as its document score.

\section{Experiments}


\subsubsection{Collection Statistics.} Table~\ref{tab:col_stat} provides details for the test collections used in our experiments.
We worked with several languages from the 2000 to 2003 Cross-Language Evaluation Forum (CLEF) evaluations~\cite{peters2001european}, using ad hoc news collections for 
French, German, Italian, Russian and Spanish.
We also conducted experiments using the new CLIR Common Crawl Collection (HC4)~\cite{hc4}, where the documents are newswire articles from Common Crawl in Chinese or Persian.
Throughout, English queries are used to search a collection in a non-English language.
We experiment with title and description queries.
The MS MARCO \cite{bajaj2018ms} passage ranking dataset, which we use for training \colbertx, consists of roughly 39M training triples, spanning over 500k queries and 8.8M passages.

\begin{table}[t]
\caption{Test collection statistics for the CLEF and HC4 newswire collections.}
\label{tab:col_stat}
\vspace{1em}
\centering
\setlength\tabcolsep{0.4em}
\begin{tabular}{@{}l|ccccccc@{}}
\toprule
\multirow{2}{*}{Collection} & HC4 & HC4 & CLEF & CLEF & CLEF & CLEF & CLEF\\
 & Chinese & Persian & French & German & Italian & Russian & Spanish \\ \midrule
\#documents & 646K & 486K & 129k & 294k & 157k & 16k & 454k \\
\#passages & 3.6M & 3.1M & 0.7M & 1.6M & 0.8M & 0.1M & 2.7M \\
\#queries & 50 & 50 & 200 & 200 & 200 & 62 & 160 \\ \bottomrule
\end{tabular}%
\end{table}

\subsubsection{\colbertx Training and Retrieval.}
Our two \colbertx model strategies, zero-shot (ZS) and translate-train (TT), are trained using mostly the same hyperparameters used to train the original ColBERT model.\footnote{We increase our batch size from 32 to 128.}
We replaced the BERT encoder with the XLM-RoBERTa (large) encoder provided by the HuggingFace transformers~\cite{wolf-etal-2020-transformers} library (but see Section~\ref{sec:ColBERT-X_models} for mBERT results).
To generate passages from documents, we use a passage length of 180 tokens with a stride of 90 tokens. We index these passages using the trained \colbertx model in the same way as the original ColBERT model in the E2E setting.\footnote{\url{https://github.com/stanford-futuredata/ColBERT#indexing}}

\subsubsection{Machine Translation.}
For CLEF languages, we use MS MARCO passage translations\footnote{\url{https://github.com/unicamp-dl/mMARCO}} from Bonifacio {\em{et al.}}~\cite{bonifacio2021mmarco}, and the same MT model to translate queries.
For the HC4 languages, we use directional MT models built on top of a transformer base architecture (6-layer encoder/decoder) using Sockeye~\cite{sockeye2}.
To produce translations of MS MARCO, the original passages were split using \textit{ersatz} \cite{wicks-post-2021-unified}, and sentence-level translation was performed using the trained MT model.

\subsubsection{Baselines.}

We compare these strategies with several baselines:
\begin{itemize}
    \item Human Translation: Monolingual retrieval using Anserini BM25~\cite{Yang2017-vd} with the document-language queries provided in the test collection. 
    \item Query Translation: BM25 retrieval using translated queries produced by a specific MT model and original documents in the target language.\footnote{To compare the retrieval models fairly,
    we use the same MT model to translate the queries as the one used to translate the MS MARCO passages.} 
    \item Reranking: We rerank query translation baseline results using the public mT5 reranker\footnote{\url{https://huggingface.co/unicamp-dl/mt5-base-multi-msmarco}}
    trained on translated MS MARCO in 8 languages~\cite{bonifacio2021mmarco}. 
    
\end{itemize}

\subsubsection{Evaluation.}
We evaluate ranking using Mean Average Precision (MAP). 
Differences in means are tested for significance using a paired t-test (p$<$0.05) with Holm-Bonferroni multiple test correction.

\begin{table*}[t!]
\caption{\label{tab:main_results} Effectiveness results (mean average precision) for CLIR HC4 and CLEF collections using title queries. Statistically significant improvements over the query translation and reranking baselines are marked with {$^*$} and ${^\dagger}$ respectively. Bold indicates best MAP among the query translation and reranking methods.}
\label{tab:my-table}
\centering
\vspace{1em}
\setlength{\tabcolsep}{0.5em}
\resizebox{\textwidth}{!}{%
\begin{tabular}{@{}l|ccccccc@{}}
\toprule
\multirow{2}{*}{\begin{tabular}[c]{@{}c@{}}Collection($\rightarrow$)\\ Model($\downarrow$)\end{tabular}} & \multirow{2}{*}{\begin{tabular}[c]{@{}c@{}}HC4\\ Chinese\end{tabular}} & \multirow{2}{*}{\begin{tabular}[c]{@{}c@{}}HC4\\ Persian\end{tabular}} & \multirow{2}{*}{\begin{tabular}[c]{@{}c@{}}CLEF\\ French\end{tabular}} & \multirow{2}{*}{\begin{tabular}[c]{@{}c@{}}CLEF\\ German\end{tabular}} & \multirow{2}{*}{\begin{tabular}[c]{@{}c@{}}CLEF\\ Italian\end{tabular}} & \multirow{2}{*}{\begin{tabular}[c]{@{}c@{}}CLEF\\ Russian\end{tabular}} & \multirow{2}{*}{\begin{tabular}[c]{@{}c@{}}CLEF\\ Spanish\end{tabular}} \\
 &  &  &  &  &  &  &  \\ \midrule
\multicolumn{8}{l}{\textit{human translation}} \\ \midrule
\multicolumn{1}{l|}{BM25} & 0.301 & 0.276 & 0.403 & 0.304 & 0.350 & 0.452 & 0.452 \\
\multicolumn{1}{l|}{\colbertx (ZS)} & 0.510 & 0.343 & 0.401 & 0.360 & 0.328 & 0.479 & 0.418 \\ \midrule
\multicolumn{8}{l}{\textit{query translation}} \\ \midrule
\multicolumn{1}{l|}{BM25} & 0.237 & 0.211 & 0.387 & 0.263 & 0.275 & 0.377 & 0.405 \\ \midrule
\multicolumn{8}{l}{\textit{reranker}} \\ \midrule
\multicolumn{1}{l|}{BM25+mT5-multi} & 0.312 & - & 0.333 & 0.297 & 0.279 & 0.303 & 0.370\\ \midrule
\multicolumn{1}{l}{\textit{our methods}} \\ \midrule
\multicolumn{1}{l|}{\colbertx (ZS)} & \textbf{0.450}$^{*\dagger}$ & 0.297{$^*$} & 0.382$^{\dagger}$ & 0.328$^{*\dagger}$ & 0.272 & \textbf{0.418}$^{\dagger}$ & 0.379\\
\multicolumn{1}{l|}{\colbertx (TT)} & 0.408$^{*\dagger}$ & \textbf{0.310}{$^*$} & \textbf{0.422}$^{\dagger}$ & \textbf{0.397}$^{*\dagger}$ & \textbf{0.339}$^{*\dagger}$ & 0.410$^{\dagger}$ & \textbf{0.415}$^{\dagger}$ \\ \bottomrule
\end{tabular}%
}
\end{table*}

\subsubsection{Results.}
Table~\ref{tab:main_results} compares the effectiveness of our models to the baselines. Our main finding is that both \colbertx variants perform better than BM25 query translation in general.
\colbertx trained using English MS MARCO alone performs better than query translation and fine-tuning \colbertx on translated MS MARCO data helps improve effectiveness further.
These gains are statistically significant in both HC4 collections, and for many CLEF collections.

We also compare the \colbertx variants to the multilingual T5 reranker that reranks the query translation baseline output. In each collection, \colbertx performs consistently and significantly better than the reranker. This is particularly interesting in CLEF collections since both the mT5 reranker and \colbertx (TT) were trained on the same MS MARCO translations. However, the reranker was trained on a combined dataset in 8 languages, which might point to the curse of multilinguality~\cite{Conneau2019-sa}.   

When we compare the two variants of \colbertx, we observe that on average translate-train often does better than zero-shot, but these differences are only significant in CLEF collections except Russian and not in HC4 collections. The difference is likely a result of using different MT models in CLEF and HC4 collections, so we conduct further analysis in the next section.

\section{Detailed Analysis}
This section considers several aspects of \colbertx. 
First, different machine translation models are compared using both MT and CLIR measures.
Second, effects of different multilingual encoders are explored.
Third, the impact of pseudo-relevance feedback is examined.
Then the influence of query length on performance is considered.
Finally, \colbertx costs in terms of index size are noted.

\subsection{Effect of Machine Translation}

\colbertx utilizes machine translation in two different ways depending on whether it is trained using the zero-shot strategy or the translate-train strategy.
In the zero-shot strategy, the queries are translated to the document language at query time, while the 
translate-train strategy requires an MT system to translate the monolingual training corpus
(in this case, the MS MARCO passages)
to the document language. The MT systems used to produce translations include:
\begin{itemize}
    \item OpusMT -- bidirectional MT model(s) with MarianNMT as the base architecture,\footnote{\url{https://huggingface.co/Helsinki-NLP}} released by the Helsinki NLP group from Bonifacio {\em{et al.}}~\cite{bonifacio2021mmarco}.
    \item SockeyeMT1 -- MT model built on top of a transformer base architecture (6-layer encoder/decoder) trained on bitext. Depending on language, these include publicly available bitext such as OpenSubtitles, UN Corpus, Europarl, and WMT. The model is trained using AWS Sockeye v2 \cite{sockeye2}. 
    \item SockeyeMT2 -- identical model to SockeyeMT1 but trained with 2x--3x more bitext. The number of training sentence pairs for MT1 vs MT2 are 51M vs 120M for Russian, 36M vs 85M for Chinese, and 6M vs 11M for Persian. 
\end{itemize}

Table \ref{tab:mt_quality} provides an intrinsic comparison of the systems translating from English on a translation task using BLEU scores~\cite{papineni-etal-2002-bleu}.
For Russian and Chinese we evaluated using a recent WMT shared task (newstest'19);
for Persian we evaluated with a collection of around 3000 sentences about COVID-19, as no WMT test is available.
Scores were calculated with \textit{sacrebleu}~\cite{post-2018-call} using the {\tt -lc} setting. 
The table reveals that SockeyeMT outputperforms OpusMT and that exposing SockeyeMT to more training data improves the BLEU score.

\begin{table}[t]
\caption{BLEU scores for translation systems using WMT'19 newstest for Chinese and Russian, and TICO-19 (from OPUS\protect\footnotemark) for Persian. These are computed on test sets distinct from the CLIR collections, so absolute BLEU score is not an exact reflection of quality of translations in CLIR experiments. Nevertheless, relative comparison of BLEU scores among MT systems is meaningful.}\label{tab:mt_quality}
\vspace{1em}
\centering
\setlength\tabcolsep{0.5em}
\begin{tabular}{@{}c|ccc@{}}
\toprule
\begin{tabular}[c]{@{}c@{}}Language\\Benchmark\end{tabular} & \begin{tabular}[c]{@{}c@{}}Russian\\ \textit{newstest'19}\end{tabular} & \begin{tabular}[c]{@{}c@{}}Chinese\\\textit{newstest'19}\end{tabular} & \begin{tabular}[c]{@{}c@{}}Persian\\\textit{tico-19}\end{tabular} \\ \midrule
OpusMT & 26.3 & 14.6 & - \\
SockeyeMT1 & 32.1 & 25.8 & 4.4 \\
SockeyeMT2 & \textbf{35.9} & \textbf{38.6} & \textbf{20.2} \\ \bottomrule
\end{tabular}%
\end{table}
\footnotetext{\url{https://opus.nlpl.eu/}}

Table~\ref{tab:mt_effect} shows that improving BLEU scores likely leads to improvements in CLIR for both training strategies.
Table~\ref{tab:mt_zs_effect} shows the results of translating queries in the zero-shot strategy. 
While BLEU improvements tend to be realized downstream, this is not seen for HC4 Chinese where OpusMT has better MAP than SockeyeMT1. 
Note that asking MT to translate keyword queries may not align well with how the systems were trained with complete sentences.

Table~\ref{tab:mt_tt_effect}
shows results for using different translation models on MS MARCO triples,
and the effect this has on \colbertx retrieval as measured using MAP.
Again, we see that the MAP scores tend to improve with improved BLEU; however, in this case the improvement in Russian BLEU from Table~\ref{tab:mt_quality}
between SockeyeMT1 and SockeyeMT2 does not carry over to \colbertx, where the performance is essentially the same.
Generally, one can expect that improving MT quality will lead to improved effectiveness of \colbertx.

\subsection{Effect of Multilingual Language Models}
\label{sec:ColBERT-X_models}
Comparing different multilingual encoders to initialize \colbertx,
we observe that \xlmr performs significantly better than mBERT, as shown in Table~\ref{tab:lm_quality}. While this might be unsurprising given that the \xlmr model is twice as large and was pretrained on more data than mBERT, tokenization differs across the languages. Considering the case of Chinese, mBERT tokenization produces character-level tokens, whereas the \xlmr tokenizer generates subwords (sentencepieces). This also implies that mBERT indexes are larger than XLM-R indexes, resulting from the term-level storage requirements of \colbertx model.

\begin{table}[t]
\caption{MAP using different MT models for \colbertx.}\label{tab:mt_effect}
\vspace{1em}
    \begin{subtable}[c]{.4\textwidth}
      \centering
        \setlength\tabcolsep{0.6em}
        \begin{tabular}{@{}c|ccc@{}}
        \toprule
        \begin{tabular}[c]{@{}c@{}}MT \\ model\end{tabular} & \begin{tabular}[c]{@{}c@{}}CLEF\\ Russian\end{tabular} & \begin{tabular}[c]{@{}c@{}}HC4\\ Chinese\end{tabular} & \begin{tabular}[c]{@{}c@{}}HC4\\ Persian\end{tabular} \\ \midrule
        OpusMT & 0.418 & 0.411 & - \\
        SockeyeMT1 & 0.442 & 0.391 & 0.230  \\
        SockeyeMT2 & \textbf{0.461} & \textbf{0.450} & \textbf{0.297} \\ \bottomrule
        \end{tabular}%
        \subcaption{\colbertx zero-shot}
        \label{tab:mt_zs_effect}
    \end{subtable}%
    \hskip5em
    \begin{subtable}[c]{.4\textwidth}
      \centering
        \setlength\tabcolsep{0.4em}
        \begin{tabular}{@{}c|ccc@{}}
        \toprule
        \begin{tabular}[c]{@{}c@{}}MT \\ model\end{tabular} & \begin{tabular}[c]{@{}c@{}}CLEF\\ Russian\end{tabular} & \begin{tabular}[c]{@{}c@{}}HC4\\ Chinese\end{tabular} & \begin{tabular}[c]{@{}c@{}}HC4\\ Persian\end{tabular} \\ \midrule
OpusMT & 0.410 & 0.365 & - \\
SockeyeMT1 & \textbf{0.459} & 0.389 & 0.287 \\
SockeyeMT2 & 0.456 & \textbf{0.408} & \textbf{0.310} \\ 
        \bottomrule
        \end{tabular}%
        \subcaption{\colbertx translate-train}
        \label{tab:mt_tt_effect}
    \end{subtable} 
    
\end{table}

\begin{table}[t]
\caption{MAP scores for \colbertx initialized with the mBERT and \xlmr encoders,
and trained on SockeyeMT1 MS MARCO translations. }\label{tab:lm_quality}
\vspace{1em}
\centering
\setlength\tabcolsep{0.6em}
\begin{tabular}{@{}c|ccc@{}}
\toprule
\begin{tabular}[c]{@{}c@{}}Multilingual\\Model\end{tabular} & \begin{tabular}[c]{@{}c@{}}CLEF\\ Russian\end{tabular} & \begin{tabular}[c]{@{}c@{}}HC4\\ Chinese\end{tabular} & \begin{tabular}[c]{@{}c@{}}HC4\\\ Persian \end{tabular} \\ \midrule
mBERT & 0.341 & 0.284 & 0.173 \\
\xlmr & \textbf{0.459}$^*$ & \textbf{0.389}$^*$ & \textbf{0.287}$^*$ \\
\bottomrule
\end{tabular}%
\end{table}

\subsection{Pseudo-Relevance Feedback} \label{sec:ColBERT-X_prf}
 
Pseudo-relevance feedback (PRF) is a form of query expansion that adds discriminative terms extracted from retrieved documents. While PRF has been explored for pre and post translation query expansion~\cite{mcnamee2002comparing}, here we choose cross-language expansion terms using the \colbertx term representation, as suggested by Wang et al~\cite{wang2021pseudo}.
First, feedback documents (fb-docs) are selected from the top of a ColBERT E2E ranked list.
Next, embeddings of terms from the feedback documents are clustered into $k$ clusters.
The top ranked centroids of these $k$ clusters\footnote{Each centroid is mapped to the nearest document token using the ANN index.}
by token IDF are used as feedback embeddings (fb-embs).
These fb-embs are added to the original query
and ColBERT E2E is run again to produce the final ranked list.
We extend this approach to the \colbertx CLIR setting. 

To better understand the effect of PRF, we compare \colbertx translate-train and query translation BM25, with and without PRF. 
For BM25, we use Anserini's RM3 to perform PRF, with default hyperparameters. 
For \colbertx PRF, we extend Terrier's~\cite{ounis2005terrier} implementation\footnote{\url{https://github.com/terrierteam/pyterrier_colbert}} with default hyperparameters. 
Table~\ref{tab:ColBERT-X_prf} shows the effect of PRF on \colbertx translate-train MAP. 
Except in Spanish, applying PRF to \colbertx significantly improves effectiveness compared to 
\colbertx without PRF or compared to BM25 with PRF.

\begin{table}[t]
\caption{MAP for query translation BM25 and \colbertx translate-train. $^*$ or $^\dagger$ denote significant improvement over BM25+PRF or \colbertx respectively}\label{tab:ColBERT-X_prf}
\vspace{1em}
\centering
\setlength{\tabcolsep}{0.6em}
\begin{tabular}{@{}l|cccc@{}}
\toprule
\begin{tabular}[c]{@{}c@{}}Retrieval \\ Model\end{tabular} &  
\begin{tabular}[c]{@{}c@{}}CLEF\\ French\end{tabular} &
\begin{tabular}[c]{@{}c@{}}CLEF\\ German\end{tabular} &
\begin{tabular}[c]{@{}c@{}}CLEF\\ Italian\end{tabular} &
\begin{tabular}[c]{@{}c@{}}CLEF\\ Spanish\end{tabular} \\
\midrule
\multicolumn{5}{l}{\textit{baseline}} \\
\midrule
BM25 & 0.387 & 0.263 & 0.275 & 0.405 \\
\colbertx & 0.422 & 0.397 & 0.339 & 0.415 \\
\midrule
\multicolumn{5}{l}{\textit{with PRF}} \\
\midrule
BM25 & 0.410 & 0.321 & 0.320 & \textbf{0.438} \\
ColBERT-X & \textbf{0.459}$^{*\dagger}$ & \textbf{0.406}$^{*\dagger}$ & \textbf{0.371}$^{*\dagger}$ & 0.436${^\dagger}$ \\ \bottomrule
\end{tabular}%
\end{table}

\subsection{Effect of Longer Queries}

Table~\ref{tab:long_queries} analyzes the effect of query type on ColBERT-X translate-train.
We compare three representations: {\em title} (t), which is a short Web-like query;
{\em description} (d), a well-formed sentence describing the information need, and {\em title+description} (td), the concatenation of the two. Longer queries pose a problem for \colbertx, however, since the model only supports queries up to 32 tokens long. To mitigate this problem, we use a list of ``stop structures''\cite{allan1998inquery} consisting of phrases (e.g., find documents on, reports of, etc.), which have been shown to work in the past, removing them from the td queries. We observe that td with stop structures removed leads to significant improvements over t or d alone.

\begin{table}[t!]
\caption{MAP results for ColBERT-X (TT) model using different query representations. $^*$ and $^\dagger$ denote significant improvements over t and d queries respectively.}\label{tab:long_queries}
\vspace{1em}
\centering
\setlength\tabcolsep{0.6em}
\begin{tabular}{c|cccc}
\toprule
\begin{tabular}[c]{@{}c@{}}Query \\ Representation\end{tabular} &  
\begin{tabular}[c]{@{}c@{}}CLEF\\ French\end{tabular} &
\begin{tabular}[c]{@{}c@{}}CLEF\\ German\end{tabular} &
\begin{tabular}[c]{@{}c@{}}CLEF\\ Italian\end{tabular} &
\begin{tabular}[c]{@{}c@{}}CLEF\\ Spanish\end{tabular}
\\ \midrule
title & 0.422 & 0.397 & 0.339 & 0.415 \\ 
description & 0.434 & 0.410 &
0.380 & 0.456  \\
title+description & \textbf{0.507}$^{*\dagger}$ & \textbf{0.466}$^{*\dagger}$ & \textbf{0.424}$^{*\dagger}$ & \textbf{0.500}$^{*\dagger}$ \\
\bottomrule
\end{tabular}%
\end{table}

\subsection{Indexing Footprint} \label{sec:ColBERT-X_footprint}

In addition to the FAISS-based ANN index,
\colbertx requires access to the representation of each term to compute MaxSim. With each term embedded as a 128-dimensional vector and each dimension using 16-bits, that's 256 bytes per term. These are onerous requirements,
with index size increasing with collection size. Table~\ref{tab:ColBERT-X_indexstats} provides statistics on storage requirements. ColBERTv2~\cite{santhanam2021colbertv2} addresses this issue by clustering token embeddings.
That approach could be extended to \colbertx for CLIR; we leave it for future work.
An artifact of our design that affects index size is how passages are generated.
We use a sliding window of document tokens, so most tokens have two representations. 
In the future, we will explore
the effects of alternative document segmentation approaches.

\begin{table}[b]
\caption{Collection-specific memory footprint.}\label{tab:ColBERT-X_indexstats}
\vspace{1em}
\centering
\setlength\tabcolsep{0.6em}
\begin{tabular}{c|ccccccc}
\toprule
\begin{tabular}[c]{@{}c@{}}Collection\end{tabular} &  \begin{tabular}[c]{@{}c@{}}HC4\\ Chinese\end{tabular} & \begin{tabular}[c]{@{}c@{}}HC4\\\ Persian \end{tabular} &
\begin{tabular}[c]{@{}c@{}}CLEF\\ French\end{tabular} &
\begin{tabular}[c]{@{}c@{}}CLEF\\ German\end{tabular} &
\begin{tabular}[c]{@{}c@{}}CLEF\\ Italian\end{tabular} &
\begin{tabular}[c]{@{}c@{}}CLEF\\ Russian\end{tabular} &
\begin{tabular}[c]{@{}c@{}}CLEF\\ Spanish\end{tabular}
\\ \midrule
\#passages & 3.6M & 3.1M & 0.7M & 1.6M & 0.8M & 0.1M & 2.7M \\
Disk Space & 154GB & 134GB & 33GB & 70GB & 36GB & 4.7GB & 117GB \\
\bottomrule
\end{tabular}%
\end{table}

\section{Conclusion}

We have developed \colbertx, a cross-language generalization of ColBERT that uses a multilingual query and document encoder to improve CLIR beyond what traditional systems such as BM25 can achieve.  Using MT systems to translate MS MARCO, we create CLIR collections for training \colbertx. We additionally analyze the effect of MT on the CLIR task.
In the future, we would like to create a single multilingual model that is trained on the data from many languages,
and compare that with a separate models for each language.
For pseudo-relevance feedback,
it is important to understand which type of queries benefit from it;
a per-query comparison could shed some light on that question.

\subsection*{Acknowledgments}

This research is based upon work supported in part by the Office of the
Director of National Intelligence (ODNI), Intelligence Advanced Research
Projects Activity (IARPA), via contract \# FA8650-17-C-9117. Views
and conclusions contained herein are those of the authors and should not
be interpreted as necessarily representing the official policies,
expressed or implied, of ODNI, IARPA, or the U.S. Government (USG). The USG is authorized to reproduce and distribute reprints for
governmental purposes notwithstanding any copyright annotation therein.

%
%
%
\bibliographystyle{splncs04}
\bibliography{bibliography}
%





\end{document}